\documentclass[aps,prl,showpacs,citeautoscript,twocolumn]{revtex4}

\usepackage{graphicx}
\usepackage{amsmath}
\usepackage{amssymb}

\newcommand{\bnen}{\begin{equation}}
\newcommand{\eden}{\end{equation}}
\newcommand{\bean}{\begin{eqnarray}}
\newcommand{\eean}{\end{eqnarray}}

\newcommand{\bna}{\begin{array}}
\newcommand{\eda}{\end{array}}

\begin{document}
\title{Graphene Andreev  Billiards}

\author{ J{\'o}zsef Cserti
\thanks{e-mail: cserti@complex.elte.hu}
}
\affiliation{Department of Physics of Complex Systems,
E{\"o}tv{\"o}s University,
H-1117 Budapest, P\'azm\'any P{\'e}ter s{\'e}t\'any 1/A, Hungary
}

\author{Imre Hagym\'asi
}
\affiliation{Department of Physics of Complex Systems,
E{\"o}tv{\"o}s University,
H-1117 Budapest, P\'azm\'any P{\'e}ter s{\'e}t\'any 1/A, Hungary
} 

\author{ Andor Korm\'anyos}
\affiliation{
Department of Physics, Lancaster University,\\
Lancaster, LA1 4YB, UK}

\begin{abstract}
We studied the energy levels of graphene based Andreev billiards 
consisting of a superconductor region on top of a monolayer graphene 
sheet.
For the case of Andreev retro-reflection we show that 
the graphene based Andreev billiard can be mapped 
to the normal metal-superconducting billiards with the same geometry. 
We also derived a semiclassical quantization rule in graphene based Andreev
billiards. 
The  exact and the semiclassically obtained spectrum 
agree very well both for the case of Andreev retro-reflection and specular
Andreev reflection.

\end{abstract}


\maketitle

In the well-known Andreev billiards consisting of a normal metal surrounded 
by a superconductor (NS) the dynamics of the quasiparticles is 
determined by the so-called 
{\it Andreev retro-reflection\/}~\cite{Andreev:cikk}. 
The spectrum of Andreev billiards is described  by the
Bogoliubov-de Gennes (BdG) equation and has been long 
studied~\cite{Kosztin:ref,Lodder-Nazarov:ref} 
(for review of the topic see, eg,~\cite{Beenakker_LectureNotes}). 

The electronic properties of graphene can be described accurately  
by massless Dirac fermion type excitations using  two dimensional relativistic quantum 
mechanics~\cite{Novoselov_graphene-1,Zhang_graphene:ref,Falko_Weak_loc_gr:ref} 
and also by semiclassical methods~\cite{carmier:245413} 
(for reviews on the physics of graphene see,
e.g.,~\cite{review_graphene:cikkek}). 
In the seminal paper by Beenakker~\cite{beenakker:067007} 
it has been shown that when monolayer graphene is interfaced 
with  a superconductor then two types of Andreev reflection are 
possible depending on the ratio of the Fermi energy 
$E_{\rm F}^{\left(\rm{G}\right)}$  and the electron energy $E$.
For $E_{\rm F}^{\left(\rm{G}\right)} > \Delta^{(G)} >E$ 
the Andreev retro-reflection is dominant as in NS billiards 
(here $\Delta^{(G)}$ is the superconducting pair potential induced in the
graphene).
In contrast, when  $E_{\rm F}^{\left(\rm{G}\right)} < E < \Delta^{(G)}$,  
 a different type of scattering process takes place   
 at the graphene-superconductor interface, which 
is named  {\it specular Andreev reflection }.  
The  specular Andreev reflection does not exist in NS systems 
and it is a prominent consequence of the peculiar band structure of 
the monolayer graphene. 
Beenakker's paper has been followed by numerous 
works~\cite{ref:beenakker-et-al}
(for a review on Andreev reflection in graphene see 
article~\cite{beenakker:1337}). 
Note that although  graphene itself is not superconducting, 
due to the proximity effect 
a superconductor can induce non-zero pair-potential in the graphene as well.
Indeed, supercurrent has been observed experimentally~\cite{Heersche_supra:ref}
 between two superconducting electrodes on top of a graphene monolayer. Moreover, 
experimental results of Ref.~\cite{ref:Miao} attest to the ballistic propagation
of quasiparticles in graphene-superconductor hybrid structures.

The most widely studied theoretical model of Andreev billiards is that of 
a two dimensional electron gas (2DEG) in a quantum dot contacted 
by a bulk superconductor
(see eg. ~\cite{Beenakker_LectureNotes}). One of the major obstacles that has 
thwarted so far the direct comparison of the theoretical predictions
and experimental results is the inevitably existing tunnel barrier 
and mismatch of the Fermi-velocities and effective masses 
between the 2DEG and the superconductor (often referred 
to as ``non-ideal NS interface'' in the literature).
This mismatch causes the probability of normal reflection to increase at the
NS interface while the probability of the Andreev reflection
diminishes significantly.
The situation when both normal and Andreev reflection 
take place at the NS interface is theoretically more difficult to address. 
In graphene however, when the superconductivity is induced by  external superconducting  
contacts,  such mismatch may not exist so that the graphene-superconductor
systems may experimentally be ideal to study most of the theoretical 
predictions made assuming perfect (ie with no mismatch) NS-interfaces.

In this paper we consider {\it graphene Andreev billiards\/} (GABs). 
In particular, we assume that in a closed region $\cal{D}$ of the
graphene sheet the superconducting pair potential is zero and outside this
region it takes on a constant value $\Delta^{(G)}$. 
We demonstrate, in one hand, that when the retro-reflection is the dominant
scattering process at the normal graphene-superconductor interface  
the electronic properties of  GABs can indeed be  obtained in  
\emph{semiclassical approximation} from the known results 
for NS billiards with ideal NS interface. 
On the other hand, we also calculate the exact spectrum of a GAB for the 
case when the dominant scattering process is the specular Andreev 
reflection and we show that it can also be understood using semiclassical
considerations.

To see the relation between the energy spectrum of NS billiards and GABs 
note the following: 
the dispersion relation of the quasiparticles 
in the normal ($\Delta^{(N)} = 0$) region 
of the NS billiards for energies $E<\Delta^{(N)}$ can be linearized 
around the Fermi energy 
$E_{\rm F}^{\left(\rm{N}\right)}$  as  
$E(p)=\pm {v}_{\rm F}^{(N)} ({ p}-p_{\rm F}^{(N)})$,
where the sign $+$ ($-$) refers to the electron-like (hole-like)
quasiparticles.
Here $p$ is the magnitude 
of the momentum, $p_{\rm F}^{(N)}= \sqrt{2m E_{\rm F}^{\left(\rm{N}\right)}}$ 
is the Fermi momentum and $v_{\rm F}^{(N)} =  p_{\rm F}^{(N)}/m$ 
is the  the Fermi velocity. 
This linearization is allowed if we are interested in the properties
of the bound states ($E<\Delta^{(N)}$) of NS  billiards  because for typical
NS billiards the dimensionless parameter $\Delta^{(N)}/E_F^{(N)}\ll 1$ 
is much less than unity. 
The \emph{same  linear dispersion} can be found for electron-like (hole-like) 
quasiparticles in the $\Delta^{(G)}=0$ region for GABs in the retro-reflection regime 
but with Fermi velocity $v_{\rm F}^{(G)}$ and 
Fermi momentum $p_{\rm F}^{(G)}$. 
This simple observation is the core of the intimate relation between the
graphene based and normal metal Andreev billiards.  
As long as the effect of the superconductor in semiclassical approximation 
can be described by  the same 
way for GABs as for the NS billiards, i.e. by
a simple phase shift $-\arccos(E/\Delta^{(N,G)})$,  one can expect that 
when the Andreev retro-reflection is the dominant scattering process 
the gross features of the energy spectrum of a GAB will closely 
resemble the spectrum of a NS billiard having the same geometry. 
This happens because the quasiparticles have linear dispersion in both cases.  
Moreover, if the Fermi velocities and Fermi momentums are the same 
i.e., $v_{\rm F} = \hbar k_{\rm F}^{(N)}/m = 
E_{\rm F}^{\left(\rm{G}\right)}/(\hbar k_{\rm F}^{(G)})$ and 
$p_{\rm F}=\sqrt{2m E_{\rm F}^{(N)}}=E_{\rm F}^{(G)}/v_{\rm F}$ 
 the  quasiparticles in the $\Delta^{(N,G)}=0$ region will  
have the same dispersion relation for both NS billiards and GABs. 
Note that if $p_{\rm F}$ and $v_{\rm F}$ are the same then 
$E_{\rm F}^{(N)}=E_{\rm F}^{(G)}/2$.

To demonstrate the idea discussed above we consider  a simple,
circular shape GAB. It consists of normal graphene region  of radius $R$ 
surrounded by superconducting graphene. 
Owing to the valley degeneracy of the Hamiltonian  the full 
BdG equation for graphene-superconductor systems 
decouples to two four by four, reduced Hamiltonians that are related 
to each other by a unitary transformation 
(see, e.g.,~\cite{beenakker:067007}). 
We now take the one corresponding to the valley ${\bf K}$. 
Due to the  circular symmetry of the setup the reduced Hamiltonian is separable 
in polar coordinates and therefore  the eigenfunctions  can be
labelled by an integer number $m$ corresponding 
to the angular momentum quantum number. 
One can show that the ansatz for the wave functions  
satisfying the Schr{\"o}dinger equation for the reduced Hamiltonian 
in the region where $\Delta^{(G)} =0$, ie, for $r<R$ with energy $E$ are 
$\Psi_{m}^{\left(\text{N}\right)}(r,\varphi) \! = 
\! \left(
c_+^{\left(\text{N}\right)}\,  
\chi^{\left(\text{N}\right)}_+(r,\varphi) 
+ c_-^{\left(\text{N}\right)} \,
\chi^{\left(\text{N}\right)}_-(r,\varphi) 
 \right) \!\! \, e^{im\varphi}$, 
where $\chi^{\left(\text{N}\right)}_+(r,\varphi) 
~= {\left[-iJ_m(k_+ r),J_{m+1}(k_+ r)e^{i\varphi},0,0\right]}^T$ and 
$\chi^{\left(\text{N}\right)}_-(r,\varphi) 
~= {\left[0,0, -iJ_m(k_- r),J_{m+1}(k_- r)e^{i\varphi}\right]}^T$ 
are the two eigenstates, and
$k_\pm ~= \left(E_{\rm F}^{\left(\rm{G}\right)} \pm E
\right)/\left(\hbar v_{\rm F} \right)$. 
In the superconducting region $r>R$ where 
the pair potential is $\Delta^{(G)}$  the wave function has the form  
$\Psi_{m}^{\left(\text{S}\right)}(r,\varphi) \!\! = 
\hspace{-2mm}\left(
c_+^{\left(\text{S}\right)}\,  
\chi^{\left(\text{S}\right)}_+(r,\varphi) 
+ c_-^{\left(\text{S}\right)} \,
\chi^{\left(\text{S}\right)}_-(r,\varphi) 
 \right) \!\! \, e^{im\varphi}$, 
where 
$\chi^{\left(\text{S}\right)}_+(r,\varphi) = 
{\left[u^{\left(\text{S}\right)}_+,v^{\left(\text{S}\right)}_+\right]}^T$,
$u^{\left(\text{S}\right)}_+= \gamma_+ v^{\left(\text{S}\right)}_+$, 
$v^{\left(\text{S}\right)}_+ = {\left[-i H_m^{(1)}(q_+ r), 
H_{m+1}^{(1)}(q_+ r)e^{i\varphi}\right]}^T$. 
The eigenstate 
$\chi^{\left(\text{S}\right)}_-(r,\varphi)$  is obtained by
the replacement $+ \to -$
and the first kind of Hankel functions to the second one and 
$q_\pm=\left(E_{\rm F}^{\left(\rm{G}\right)} 
\pm i\sqrt{[\Delta^{(G)}]^2-E^2}\right)/(\hbar v_{\rm F})$, 
while $\gamma_{\pm} = e^{\pm i \arccos \left(E/\Delta^{(G)}\right)} $.
Here $J_{m}(x) $ and $H_{m}^{\left(1,2 \right)}(x)$ are the Bessel and
the Hankel functions~\cite{Abramowitz:book}. 
To ensure that the wave function of the bound states is normalizable, 
the wave function in the superconducting region 
must go to zero as $r \to \infty$. 
This condition can be satisfied by choosing the appropriate Hankel function 
in the eigenstates 
$\chi^{\left(\text{S}\right)}_\pm(r,\varphi)$~\cite{Abramowitz:book}. 
Finally, the unknown coefficients $c_{\pm}^{\left(\text{N}\right)}$ and 
$c_{\pm}^{\left(\text{S}\right)}$ can be determined 
from the boundary conditions
$\Psi_{m}^{\left(\text{N}\right)}(r=R,\varphi) =
\Psi_{m}^{\left(\text{S}\right)}(r=R,\varphi)$ valid for any $\varphi$. 
Thus, the condition for non-trivial solutions of the coefficients 
$c_{\pm}^{\left(\text{N}\right)}$ and $c_{\pm}^{\left(\text{S}\right)}$ 
can be found from the zeros of a four by four determinant. 
After some algebra we obtain a quite simple secular equation for the energy
levels with fixed angular momentum index $m$: 
\begin{subequations}
\label{szek0:eq}
\begin{eqnarray}
\label{DGS:eq}  
&&\hspace{-10mm} \rm{Im} \left \{\gamma_+ D^{\rm{(+)}}_{\rm{GS}}(m,E) \, 
D^{\rm{(-)}}_{\rm{GS}}(m,E) \right \} = 0, \\[1ex]
\label{De-disk_GS:eq} 
D^{\rm{(+)}}_{\rm{GS}}(m,E) &=&  
\left| \begin{array}{cc}
J_m(k_+ R)  & H_m^{(1)}(q_+ R)\\
J_{m+1}(k_+ R)  & H_{m+1}^{(1)}(q_+ R)
\end{array} \right|, 
\end{eqnarray}
\end{subequations}
and $D^{\rm{(-)}}_{\rm{GS}}(m,E) =
{\left[D^{\rm{(+)}}_{\rm{GS}}(m,-E)\right]}^*$, 
and  ${\rm Im}\{. \}$  and $* $ stand for the imaginary part and 
the complex conjugation, respectively.
Note that $H_m^{(2)}(q_- R) = {\left[H_m^{(1)}(q_+ R)\right]}^*$.
The solutions of Eq.~(\ref{szek0:eq})  for $m=0,\pm 1,\pm 2,\cdots,$ 
are the exact energy levels of a circular shape GAB. Note that 
Eq.~(\ref{szek0:eq}) is  valid both in the case of 
Andreev retro-reflection 
($E_{\rm F}^{\left(\rm{G}\right)} > \Delta^{(G)} > E$) and for specular 
Andreev reflection ($E_{\rm F}^{\left(\rm{G}\right)} < E < \Delta^{(G)}$). 
One can also notice that the eigenenergies  depend only on two dimensionless
parameters: $E_{\rm F}^{\left(\rm{G}\right)}/\Delta^{(G)} $ 
and $\xi_c^{(G)}/R$, where $\xi_c^{(G)}=\hbar v_{\rm F}^{(G)}/\Delta^{(G)}$ 
is the coherence length in the superconducting graphene.

\begin{figure}[hbt]
\includegraphics[scale=0.65]{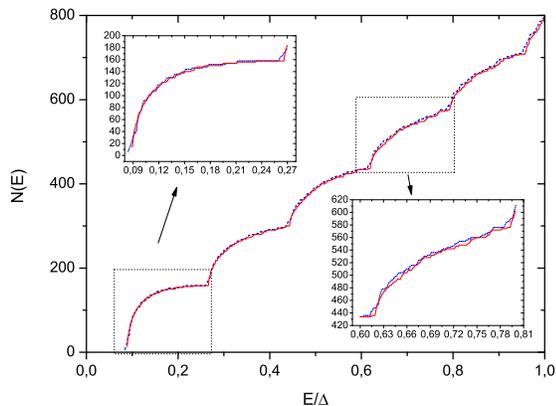}
\caption{\label{lpecso-retro:fig}  
(color online) 
The exact step function $N(E)$ for circular shape GAB (red line) and NS (blue line) 
billiards in the case of Andreev retro-reflection. 
The parameters for GAB and NS billiard are 
$E_{\rm F}^{\left(\rm{G}\right)}/\Delta^{(G)} = 10$ and 
$E_{\rm F}^{\left(\rm{N}\right)}/\Delta^{(N)} = 5$,
respectively, and $\xi_c^{(G)}/R = 0.12$ for both cases. 
The insets show the enlarged parts of the main frame.
The trivial factor 2 owing to the valley degeneracy in graphene is not
included.}
\end{figure}

We now compare the density of states (DOS) $\varrho (E) =
\sum_{nm}\delta(E-E_{nm})$  of a circular shape GAB and of the
corresponding NS billiard. 
For details of the calculation see a similar calculation for 
NS billiards in Ref.~\cite{sajat_cake:cikk}. 
It is more convenient to plot the integrated DOS, namely the so-called step
function $N(E)= \sum_{nm}\Theta(E-E_{nm})$, 
where $\Theta(x)$ is the Heaviside  function. 
Our numerical results for 
$E_{\rm F}^{\left(\rm{N}\right)}/\Delta^{(N)} 
= E_{\rm F}^{\left(\rm{G}\right)}/(2\Delta^{(G)})$ 
are shown in Fig.~\ref{lpecso-retro:fig}. 
One can see that  the step functions for the considered 
 NS billiard and GAB are indeed very similar. 
It is also clear from Fig.~\ref{lpecso-retro:fig} that the DOS
$\varrho (E) =dN(E)/dE$ shows singularities at certain energies 
$E_n^{\left(\rm sing \right)}$.
Singularities of this kind arise in the case of NS billiards as well 
(see e.g. Ref.~\cite{sajat_cake:cikk}) and 
we shall discuss their origin below. 

We now demonstrate  on the example of circular GABs that 
at semiclassical level the results for NS billiards and GABs 
can  be mapped into each other by choosing the parameters appropriately. 
 In numerous works~\cite{Melsen,Lodder-Nazarov:ref,Heny,Ihra-Richter1:ref,ref:adagideli,sajat_cikkek}
it was shown that for NS billiards in semiclassical approximation 
the step function reads
\begin{subequations}
\label{BS_Stepfn:eq}
\begin{eqnarray}
N_{\rm BS} (E) &=& M \,\sum_{n=0}^\infty \,\left\{1-F[s_n(E)]\right\}, 
\label{F_s:eq} \\
s_n(E) &=& \frac{n\pi +\arccos \left(E/\Delta^{(N)} \right)}{E/\Delta^{(N)}} \,
\xi_c^{(N)}. \label{s_n:eq} 
\end{eqnarray}
\end{subequations} 
Here $M$ is the number of open channels in the normal region, 
 $\xi_c^{(N)}=\hbar v_{\rm F}^{(N)}/\Delta^{(N)}$ 
is the coherence length in the NS system, 
$F(s) = \int_0^s \, P(s^\prime) \, ds^\prime$ is the integrated path
length distribution and $P(s)$ is the classical probability
that an electron entering the billiard at the NS interface
returns to the interface after a path of length $s$.  
The path length distribution $P(s)$ is normalized to
one, i.e., $\int_0^\infty\, P(s)\, ds =~1$ and 
one can see that it is a purely geometry-dependent function. 
In particular, for circular billiards 
it was found that 
$P(s) = \frac{1}{{\left( 2R \right)}^2} \, 
\frac{s}{\sqrt{1-{\left(s/2R\right)}^2}} \, \Theta(2R-s)$ 
and $M=2\pi\, k_{\rm F}^{(N)} R$~\cite{sajat_cake:cikk}. 
Finally, the quantity $s_n(E)$ in Eq.~(\ref{s_n:eq}) 
depends on the quantization
condition for the periodic motion of the electron-hole 
quasiparicles~\cite{Melsen,ref:adagideli}.
As it has been pointed out in the introduction, 
in good approximation the quasiparticles 
have linear dispersion in the non-superconducting region
for both GABs and NS billiards. 
If the effect of the superconductor in GABs can be taken into account 
by a simple phase shift 
$-\arccos(E/\Delta^{(G)})$~\cite{beenakker:1337},
expressions of the type of Eq.~(\ref{BS_Stepfn:eq}) can be used 
to calculate the semiclassical approximation of $N(E)$ for GABs as well.

Moreover, employing the same steps as in Ref.~\cite{sajat_cake:cikk}, 
from Eq.~(\ref{szek0:eq}) 
one can derive the following semiclassical quantization rule 
for circular shape GABs: 
\begin{subequations}
\label{BS_E_m:eq}
\begin{eqnarray}
\label{rad_action_quant:eq}
&& \hspace{-7mm} S_+(E) \! - \! \mu_r \, S_-(E) 
 \! - \! 2\arccos \frac{E}{\Delta^{(G)}} 
= 2 \pi \! \left(\!\! n+\frac{1-\mu_r}{4}\!\! \right),  \\[2ex]
&& \hspace{-7mm} 
S_\pm (E) = 2\sqrt{{\left(\left|k_\pm \right| R\right)}^2-m^2} 
-2\left|m \right| \arccos \frac{\left| m \right|}{\left|k_\pm \right| R},
\label{semi_quant:eq} 
\end{eqnarray}
\end{subequations}
where $\mu_r = 1, -1$ for Andreev retro-reflection 
and specular Andreev reflection, respectively, and 
$n$ is a non-negative integer. 
Functions $ S_\pm (E)$ are the radial action (in units of $\hbar$) of 
electrons and holes~\cite{Brack:book} and 
the term $-2\arccos E/\Delta^{(G)}$ in Eq.~(\ref{rad_action_quant:eq}) accounts
for the two Andreev reflections in one period of the orbit, while 
the second term in the left hand side of Eq.~(\ref{rad_action_quant:eq}) 
results from the sum and the difference of the Maslov indices $\pi/4$ 
of the electron-like and hole-like particles 
for $\mu_r = 1$ and $\mu_r = -1$, respectively.

\begin{figure}[hbt]
\includegraphics[scale=0.65]{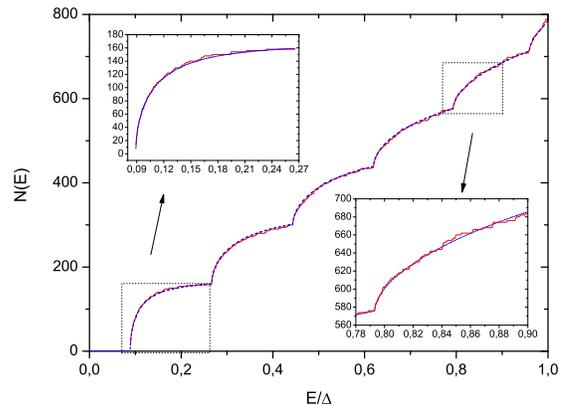}
\caption{\label{lpecso-semi-retro:fig}  
(color online)
The exact (red line) and the semiclassically (blue line) calculated 
step function $N(E)$ obtained from Eqs.~(\ref{szek0:eq}) 
and (\ref{BS_Stepfn:eq}), respectively 
for the case of Andreev retro-reflection.  
The parameters are the same as in Fig.~(\ref{lpecso-retro:fig}). 
The insets show the enlarged parts of the main frame.
}
\end{figure}

Formally, in the case of  Andreev retro-reflection the quantization condition shown 
in Eq.~(\ref{BS_E_m:eq}) 
is the same as for a circular NS billiard~\cite{sajat_cake:cikk}, 
but the meaning of $k_{\pm}$ is different for the two systems (for 
 NS billiards see eg. Ref.~\cite{sajat_cake:cikk}). 
However, from Eq.~(\ref{BS_E_m:eq})  it is easy to find that if 
$R^{(N)}/\xi_c^{(N)}=R^{(G)}/\xi_c^{(G)}$  
and $E_F^{(N)}/\Delta^{(N)}=E_F^{(G)}/(2 \Delta^{(G)})$ 
then to first order in  $E/\Delta^{(N,G)}$ the 
quantization condition for circular GABs and NS billiards is the same 
and the step function $N(E)$ is given by Eq.~(\ref{BS_Stepfn:eq}) 
with coherence length $\xi_c^{(G)}$. 
The exact and semiclassically calculated $N(E)$ are 
plotted in Fig.~\ref{lpecso-semi-retro:fig}.
The agreement between the two results  is excellent.  
Moreover, from Eq.~(\ref{BS_Stepfn:eq}), 
we find that the positions of 
the singularities in the DOS are given by
$E_n^{\left(\rm sing \right)}/\Delta^{(G)} = \left(n+1/2 \right)\pi /(1+ 2R/\xi_c^{(G)})$ 
valid for such integers $n$  that 
$E_n^{\left(\rm sing \right)} < \Delta^{(G)}$ holds.
Note that the position $E_n^{\left(\rm sing \right)}/\Delta^{(G)}$ 
of the singularities depends only on $R/\xi_c^{(G)}$ but not on 
$E_{\rm F}^{(G)}/{\Delta^{(G)}}$. 
Therefore even if 
$E_{\rm F}^{(N)}/\Delta^{(N)}\neq E_{\rm F}^{(G)}/(2 \Delta^{(G)})$ 
but  $R^{(N)}/\xi_c^{(N)}=R^{(G)}/\xi_c^{(G)}$, the singularities 
in the DOS for a circular  GAB and NS billiard  would appear 
at the same energies. 

\begin{figure}[hbt]
\includegraphics[scale=0.65]{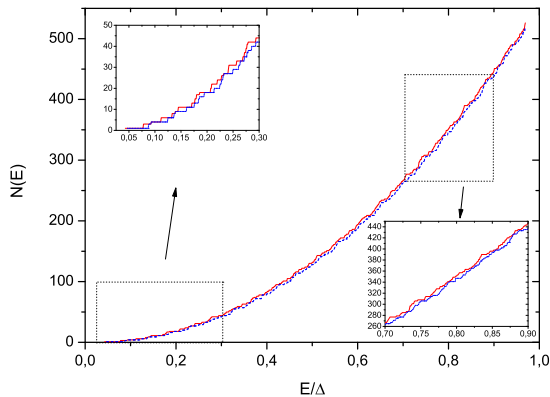}
\caption{\label{lpecso-spec:fig}  
(color online) 
The exact (red line) and the semiclassically (blue line) calculated 
step function $N(E)$ obtained from (\ref{szek0:eq}) 
and (\ref{BS_E_m:eq}), respectively for specular Andreev reflection.  
The parameters are $E_{\rm F}^{\left(\rm{G}\right)} = 0$ 
and $\xi_c^{(G)}/R= 0.03$. 
The insets show the enlarged parts of the main frame.
}
\end{figure}

Next we  consider the case of specular Andreev reflection in graphene 
Andreev billiards. 
Again, the solutions of Eqs.~(\ref{szek0:eq}) and (\ref{BS_E_m:eq})
give the exact and the semiclassically calculated energy levels of 
circular shape GABs.
In Fig.~\ref{lpecso-spec:fig} the calculated step function $N(E)$ is plotted and
as one can see it is completely different from that obtained for
the case of Andreev retro-reflection shown in Fig.~\ref{lpecso-retro:fig}.
Moreover, the quantum results in Fig.~\ref{lpecso-spec:fig} again show very good 
agreement with the semiclassical ones that can be obtained from 
Eq.~(\ref{BS_E_m:eq}). 
This implies  that in the case of 
specular Andreev reflection the DOS depends linearly on the energy for $E\rightarrow 0$. 
Namely, it can be shown from Eq.~(\ref{BS_E_m:eq}) 
that in this limit 
the DOS in semiclassical approximation (without the valley degeneracy) is given by  
$\rho(E) = 8\frac{E \mathcal{A}}{\pi^3(\hbar v_F)^2} $,   
where $\mathcal{A}$ is the area of the billiard. It is interesting to
note therefore that   (apart from the valley degeneracy) 
$\rho(E)$ is bigger by a factor of $16/\pi^2$ than in the
case of neutrino billiards~\cite{ref:berry}. 

In summary, we calculated the energy levels of graphene based Andreev
billiards. We showed that for energy levels corresponding to the case
of Andreev retro-reflection the graphene based Andreev
billiards in a very good approximation can be mapped to 
the normal metal-superconducting billiards with the same geometry. 
We also derived a semiclassical quantization rule in graphene based Andreev
billiards and the spectrum obtained from this rule agrees 
very well with that obtained from the exact quantum calculations 
for circular shape of GS billiards.

This work is supported by the Hungarian Science Foundation OTKA 
under the contracts No. T48782 and 75529.
A. K. is supported partly by European Commission Contract
No.~MRTN-CT-2003-504574 and by EPSRC.
We gratefully acknowledge fruitful discussions 
with  C. W. J. Beenakker and V. Falko.

\end{document}